\documentclass[10pt,twocolumn,letterpaper]{article}

\usepackage{wacv}
\usepackage{times}
\usepackage{epsfig}
\usepackage{graphicx}
\usepackage{amsmath}
\usepackage{amssymb}
\usepackage{booktabs}
 
\usepackage{subfig}
\usepackage{CJKutf8}
\usepackage[russian,english]{babel}

\newcommand{\SystemName}{RWDF-23}

\wacvapplicationstrack 

\wacvfinalcopy 

\ifwacvfinal
\usepackage[breaklinks=true,bookmarks=false]{hyperref}
\else
\usepackage[pagebackref=true,breaklinks=true,colorlinks,bookmarks=false]{hyperref}
\fi

\begin{document}

\title{Towards Understanding of Deepfake Videos in the Wild}

\author{Beomsang Cho\thanks{Both authors contributed equally to this research.}\\
{\tt\small gababsang@g.skku.edu}\\
Sungkyunkwan University\\ South Korea\\
\and
Binh M. Le\footnotemark[1]\\
{\tt\small bmle@g.skku.edu}\\
Sungkyunkwan University\\ South Korea\\
\and
Jiwon Kim\\
{\tt\small merwl0@g.skku.edu}\\
Sungkyunkwan University\\ South Korea\\
\and
Simon Woo\thanks{Corresponding author.}\\
{\tt\small swoo@g.skku.edu}\\
Sungkyunkwan University\\ South Korea\\
\and
Shahroz Tariq\\
{\tt\small shahroz.tariq@data61.csiro.au}\\
CSIRO's Data61\\ Australia
\and
Alsharif Abuadbba\\
{\tt\small sharif.abuadbba@data61.csiro.au}\\
CSIRO's Data61\\ Australia
\and
Kristen Moore\\
{\tt\small kristen.moore@data61.csiro.au}\\
CSIRO's Data61\\ Australia
}

\maketitle
\thispagestyle{empty}

\begin{abstract}
     Deepfakes have become a growing concern in recent years, prompting researchers to develop benchmark datasets and detection algorithms to tackle the issue. However, existing datasets suffer from significant drawbacks that hamper their effectiveness. Notably, these datasets fail to encompass the latest deepfake videos produced by state-of-the-art methods that are being shared across various platforms. This limitation impedes the ability to keep pace with the rapid evolution of generative AI techniques employed in real-world deepfake production. Our contributions in this {IRB-approved} study are to bridge this knowledge gap from current real-world deepfakes by providing in-depth analysis. We first present the largest and most diverse and recent deepfake dataset (\SystemName) collected from the wild to date, consisting of 2,000 deepfake videos collected from 4 platforms targeting 4 different languages span created from 21 countries: Reddit, YouTube, TikTok, and Bilibili. By expanding the dataset's scope beyond the previous research, we capture a broader range of real-world deepfake content, reflecting the ever-evolving landscape of online platforms. Also, we conduct a comprehensive analysis encompassing various aspects of deepfakes, including creators, manipulation strategies, purposes, and real-world content production methods. This allows us to gain valuable insights into the nuances and characteristics of deepfakes in different contexts. Lastly, in addition to the video content, we also collect viewer comments and interactions, enabling us to explore the engagements of internet users with deepfake content. By considering this rich contextual information, we aim to provide a holistic understanding of the {evolving} deepfake phenomenon and its impact on online platforms. 
\end{abstract}


\section{Introduction}





In recent years, the emergence of deepfakes in the form of videos and images has provoked critical concerns about their potential misuse \cite{romano2019, patrini2019,quandt2019fake, fbi2022,pentagon,tariq2022real,tariq2023evaluating,Metaverse}, {ranging from pornographic and harmful videos to fake information generation.} 
Notably, recent fake images on the fire near Pentagon have created the turmoil for the stock market~\cite{pentagon}, which shows {how deepfakes can affect society, and the economy with their fake information.} Meanwhile, there has been significant progress in defending against this deepfake content~\cite{li2018exposing, sebyakin2021spatio, zhao2023proactive,CLRNet,CoRed,FReTAL,PTD,TAR,lee2021detecting,khalid2021evaluation,CLRNetold,ShallowNet1,ShallowNet2, woo2022add, woo2023qad, binh2021exploring,le2023deepfake}. Concurrently, substantial efforts have also been dedicated to introducing state-of-the-art (SoTA) deepfake datasets to support researchers worldwide who are engaged in studying this urgent issue.

An influential dataset known as FaceForensics++ was introduced by R\"ossler \emph{et al.}~\cite{Rossler2019ICCV}. This dataset employs four distinct techniques, spanning from face reenactment to face swapping methods, generating 1,000 videos for each approach. Consequently, various other datasets have also been introduced, each encompassing a range of methods and diversity in subjects featured in the videos  \cite{Celeb_DF_cvpr20,  dolhansky2020deepfake, jiang2020deeperforensics, kwon2021kodf, Zhou_2021_CVPR, le2021openforensics,thakral2023phygitalnet}. 
However, these works focus primarily on producing benchmark datasets, while lacking a comprehensive understanding of the types of videos present in the real world, the platforms where deepfake videos are prevalent, and the distinctive characteristics of deepfake production in various countries. 

 In another line of endeavor, several works have put their efforts into collecting deepfake datasets from real-world environments. {In 2021,} 
 Pu \textit{et al.}~\cite{pu2021deepfake} published a real-world dataset called DF-W, composed of 1,869 videos sourced from platforms such as YouTube and Bilibili. Around the same time, Zi \textit{et al.} \cite{zi2020wilddeepfake} introduced another collection, the WildDeepfake dataset, which comprises 707 videos obtained from various internet sources. 
 However, due to the constant emergence of new technologies and sophisticated videos each year, their datasets lack up-to-date content. Additionally, these datasets exhibit a limited diversity of attributes, creators, platforms, and viewers, which creates barriers to performing more in-depth analysis of questions such as {who, where, when, and how deepfakes are created and shared in the real world. In particular, it is already well-known that deepfake videos are maliciously used for producing pornographic materials~\cite{deepfakeporn}. How are the real-world deepfakes used for other than lewd cases?

 
 To address these challenges, we have initiated the collection of the largest and most diverse dataset to date. 
Our study aims to broaden our understanding of real-world deepfake datasets, investigate the latest misuse cases, and delve into the creation and sharing phenomena of deepfakes across 21 countries, investigating 4 platforms with 4 different languages. In this study, we focus on the analysis of non-lewd video content, an area that has received limited research attention thus far. Through an IRB-approved study, we seek to answer the following research questions:

$\bullet$ \textbf{RQ1. } What are the characteristics of recently generated deepfake videos shared online over different platforms across countries? 

$\bullet$  \textbf{RQ2. } Who are the deepfake video uploaders, and what are their intentions? 

$\bullet$  \textbf{RQ3. } How do users' responses and reactions to deepfake videos change over time?

In order to answer the above research questions, we first systematically collect the latest deepfake videos and created a dataset, called ~\SystemName, of 2,000 videos from the wild {over 4 different platforms}. In addition to videos, we also collect the associated metadata and attribute information such as creators/uploaders, country, subjects' characteristics in the videos (gender, ethnicity, profession, etc), as well as viewer interaction (view, like, comment, and timestamp) to improve the understanding of real-world deepfake creation and purpose. Next, we embark upon the task of comparing and analyzing its diverse range of deepfakes, and further categorize and classify the video uploaders and contents. In addition, with interactive user responses and reactions, we obtain valuable insights that can enable the development of effective strategies to address real-world deepfake challenges. In particular, we leverage NLP tools to analyze video comments to assess the audience's impression of the deepfakes' realism and how factors such ethnicity and gender correlate with audience sentiment serving as identifying the different forms of malicious exploitation in the future.

Our main contributions are as follows: 1) We provide \textbf{the most comprehensive and diverse up-to-date dataset} of deepfake videos collected from real-world sources from YouTube, TikTok, BilliBilli, and Reddit over querying 4 different languages created from 21 different countries, reflecting the latest deepfake generation methods. 2) We conduct \textbf{in-depth empirical analysis and characterization} on how non-pornographic real-world deepfakes are created, used, and shared over different platforms, countries, races, and genders for different purposes. And, 3) We capture and analyze \textbf{user responses and reactions to deepfakes} from multiple perspectives, considering factors such as audience impression, sentiment, and the impact of deepfakes on trust and perception.

 \begin{figure}[!t]
    \centering
    \includegraphics[width=7.8cm]{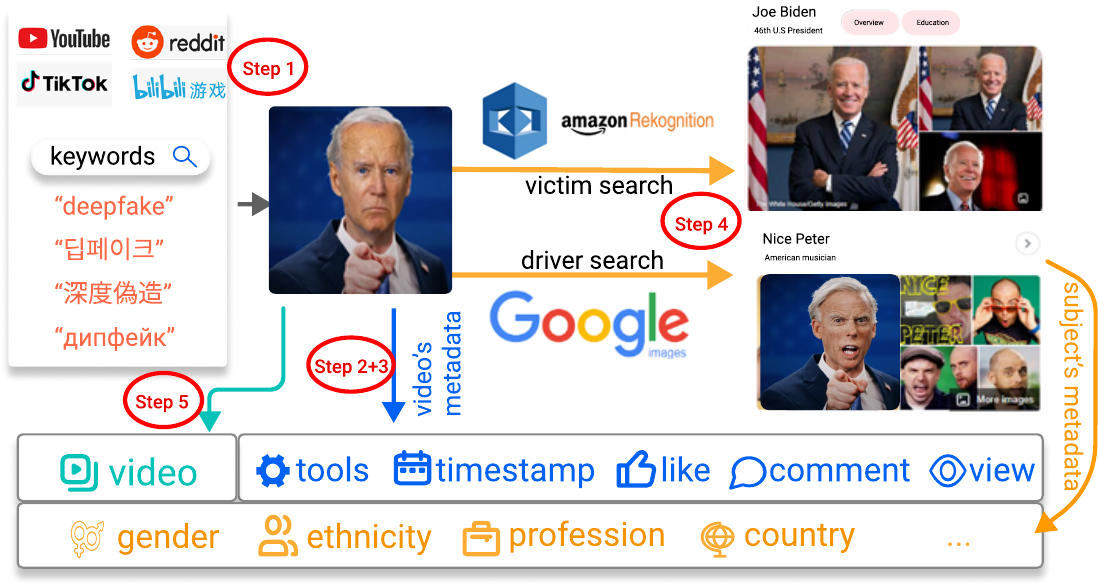}
    \caption{Summary of our overall procedure to collect the dataset. Besides the videos, we have collected their metadata and also labeled various attributes of the content.} 
    \label{fig:overall}
\end{figure}

\vspace{-2pt}
\section{Data collection methodology}
We present our newly curated dataset of deepfake videos, {named} ~\SystemName, which stands for \textit{Real-World Deepfake} in \textit{2023}, which was collected during the period of time from 2023-01 to 2023-05.
It encompasses a wide range of deepfake videos sourced exclusively from four distinct online media-sharing platforms, targeting four different languages.
We first introduce more details on various attributes we plan to collect along with videos.

\begin{table*}[!t]
\centering
\caption{Summary of \SystemName\ dataset.}
\label{tab:sum}
\resizebox{0.85\textwidth}{!}
    {%
    \begin{tabular}{l|c|c|c|c|c|c|c|c} 
    \toprule
    Dataset  & \# Videos & \# Uniq Accounts & \# Uniq Used Faces & \# Likes   & \# Comments & \# Views    & Total Duration & Avg. Duration \\ \hline
    \hline
    YouTube  & 1,190     & 107              & 616                & 3,187,806  & 264,270     & 187,059,269 & 39hr48m30s     & 2m2s \\ \hline
    TikTok   & 599       & 84              & 260                & 41,357,828 & 550,348     & No           & 3hr4m32s      & 18s  \\ \hline
    Reddit   & 111       & 14               & 72                 & 2,356      & 366         & No           & 3hr24m11s      & 1m50s\\ \hline
    Bilibili & 100       & 53               & 71                 & 22,101     & 2,094       & 1,557,440   & 1hr51m39s      & 1m8s \\ \hline
    Total    & 2,000     & 258              & 1,019                & 44,570,091 & 817,078     & 188,616,709 & 48hr8m52s     & 1m26s \\ 
    \bottomrule
    \end{tabular}%
    }
\end{table*}
\noindent \textbf{Platforms. } 
We seek to understand which platforms deepfake creators/uploaders and consumers typically share and post their deepfake videos to. Also, our aim is to investigate the unique characteristics of each platform associated with deepfakes, which includes analyzing the types of content shared by users and investigating the patterns of interactions within these platforms.

\noindent \textbf{Intents/Purposes.} 
We aim to investigate the underlying motivations behind the creation and dissemination of deepfakes by uploaders, shedding light on the factors driving their popularity and impact. By carefully curating a diverse range of deepfake content and classifying them into different categories (entertainment, political, fraud, or other purposes), we strive to uncover reasons behind the increasing enthusiasm and popularity of deepfake videos in the online community. Also, we seek to examine the societal implications, ethical considerations, and potential risks associated with the widespread adoption and consumption of deepfake content. 

\noindent \textbf{Creators/Uploaders. }We seek to examine and characterize the uploaders, 
including which countries they predominantly represent from, and further what content category they mainly create. We explore the demographics of deepfake distributors, including patterns that emerge between geographical locations and the deepfake content they share, by analyzing available uploader metadata.
This enables a better understanding of entity groups that are predominantly involved in deepfake production. On top of that, we probe into the country-specific characteristics, such as prevalent deepfake methods, popular targets, and cultural factors that may influence the creation and dissemination of deepfakes. 

\noindent \textbf{Generation Methods. } We are also interested in capturing the current status of specific tools that are popularly used in real-world deepfake creation. These intricate generation methods play a vital role in the creation of deepfake videos, and gaining a comprehensive understanding of them is essential for improving the detection accuracy
~\cite{mirsky2021creation}. For videos that state their creation method, we analyze the specific algorithms employed in deepfake production (such asFaceSwap), which reveals the key challenges and vulnerabilities associated with detecting and mitigating this emerging threat. 


\noindent \textbf{Timing and Trends. } We examine whether deepfake generation changes significantly over time, or in response to other technical or societal events. This investigation is important in understanding the landscape of deepfakes and predicting potential turning points in the future under the influence of uploaders' and viewer responses' behavior. This data enables us to understand the temporal aspects of deepfake dissemination, including the timeline of its emergence, potential shifts in popularity over time, and any correlations with significant events or developments in the field.


\noindent \textbf{User Interactions. }The next interesting question is to observe how deepfake videos receive comments and interactions from real-world users. We strive to uncover the extent of their influence and the implications they have on various aspects of society. 
We conduct a user interaction study based on these guidelines, examining the number of likes on deepfake videos to determine which videos generate enthusiasm, investigating comments to understand how people react to deepfake videos, and we measure the sentiment of these comments to identify trends in popular deepfake video content. This investigation illuminates the ways in which deepfakes permeate and interact with our daily lives, impacting areas such as politics, entertainment, misinformation, and public perception.

\subsection{\SystemName\ Collection Overview}

We present the overall data collection pipeline in Fig.~\ref{fig:overall}, where more detailed steps are explained below.

\noindent \textbf{Step 1. } 
We select the most popular platforms based on the user base, number of uploaders, level of activity, and geographical popularity to prevent data bias toward specific countries. As a result, four popular platforms are selected: YouTube, TikTok, Reddit, and Bilibili. For each platform, to identify deepfake videos, we include only those videos that are clearly labeled as deepfakes in their titles, querying the following keyword in four different languages (English, Korean, Chinese, Russian):  
``Deepfake'', \begin{CJK}{UTF8}{mj}
``딥페이크'',
\end{CJK}
\begin{CJK*}{UTF8}{bsmi}
``深度偽造'', 
\end{CJK*} and \foreignlanguage{russian}{``дипфейк''}. 
During the first data collection, we identify re-posted URLs, representing duplicate deepfake videos (144 on YouTube, 5 on TikTok, and 1 on Reddit). After filtering, we acquire 2,000 distinct URLs from 258 channels: 1,190 from YouTube, 599 from TikTok, 111 from Reddit, and 100 from BiliBili.

\noindent \textbf{Step 2. } 
 Besides the deepfake video itself, we gather pertinent information about the uploader.  We collect the uploader's profile, past uploads, and any available public information to gain insights into their background and affiliation. {Consequently, later analysis can provide us with a comprehensive understanding }of the countries from which deepfakes predominantly originate, and any distinctive regional characteristics. Furthermore, we expand our data collection by investigating additional deepfake videos uploaded by the identified uploader, {enabling a more comprehensive dataset and deeper analysis.} 
 Particularly, we examine various aspects such as the {uploading timestamp}, viewer engagement metrics (likes, comments, and views), and video duration to capture a holistic understanding of the deepfake videos and user responses to the deepfake videos. Our final collection spans 21 countries with 258 channels. 
 
\noindent \textbf{Step 3. } 
It is important to understand the underlying technologies used in deepfake creation \cite{nadimpalli2022improving}. Therefore, we collect detailed information about the tools employed to produce deepfake content. 
We take the following approaches to ensure accurate information as follows: We either gather the information from the publisher's post, directly inquire with them, or examine the embedded watermark in the video. 
For instance, in the video ``Keanu Reeves as Morpheus [DeepFake]'', DeepFaceLab~\cite{DeepFaceLab} is mentioned along with a link to its GitHub repository. 
Through these approaches, we identify information about deepfake generation methods or applications, totaling 465 videos. This accounts for approximately 23.3\% of our 2,000 videos. As shown in Fig.~\ref{fig:prop_model}, a total of 16 generation methods are identified and 75\% of the videos were created by DeepFaceLab.

\noindent \textbf{Step 4 \& 5.} As a crucial step, we collect the identity of the facial subject (\textbf{victim}) in deepfake videos and the person portrayed in the original video (\textbf{driver}).
If their names are provided in the video description, we use the popular search engine to gain further information about the victim's occupation, gender, country of birth, race, and skin color. In cases where information is not present in the video description, we extract the victim's face and apply Amazon Rekognition's Celebrity API\footnote{\url{https://aws.amazon.com/rekognition/}} to recognize them. For victims with a confidence level above 90\%, we utilize the API-provided information. Those with lower confidence levels are labeled as unknown. Similarly, for the driver, whose face has changed but the background remains the same, we capture the scene with the background and conduct Google Image\footnote{\url{https://images.google.com/}} to locate the original video with the same background. If found, we employ the Amazon Rekognition API to identify the driver's identity. In Fig.~\ref{fig:overall}, for example, we determine the subject as Joe Biden and associate character traits from the original video with Nice Peter.



The overall dataset statistics are presented in Table \ref{tab:sum}, summarizing the essential information of collected videos. As shown, our dataset is more comprehensive with respect to the number of platforms, metadata, and attributes, including more recent and diverse deepfake videos, compared to DF-W and WildDeepfake. 

\begin{figure*}[!ht]
\vspace{-10pt}
    \centering
        \subfloat[Growth across platforms 
    ]{%
        \raisebox{-3.85cm}{\includegraphics[width=3.8cm]{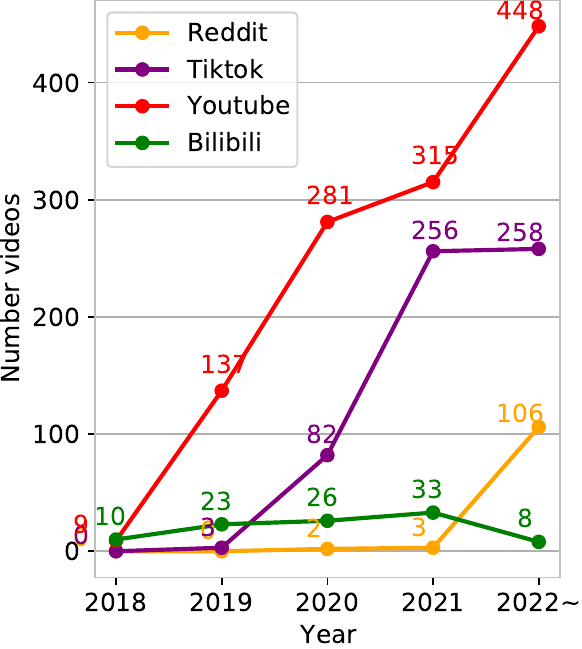}}
        \label{fig:when}%
    }\hspace{18pt}
    \subfloat[Proportion of deepfake applications ]{%
        \raisebox{-3.9cm}{\includegraphics[width=4.8cm]{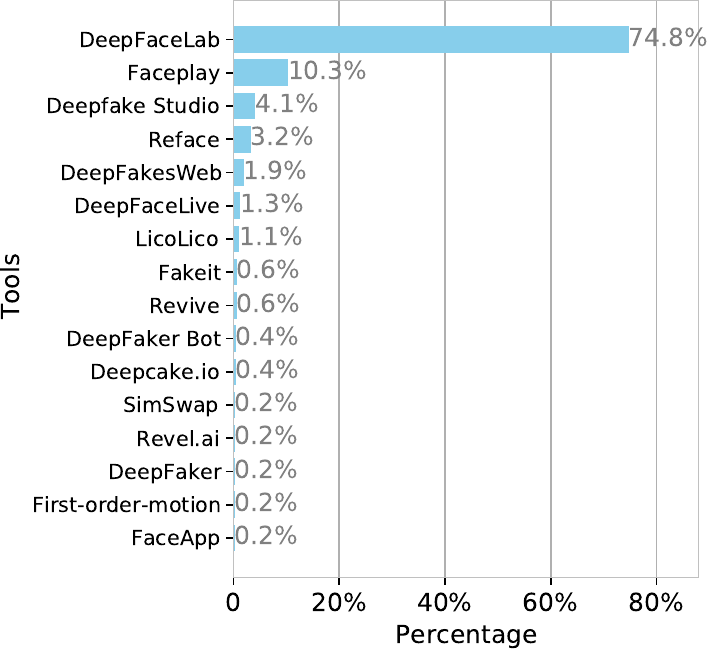}}
        \label{fig:prop_model}%
    }\hspace{18pt}
    \subfloat[Geographic distribution of publisher ]{%
        \raisebox{-3.9cm}{\includegraphics[width=5.4cm, height=4.4cm]{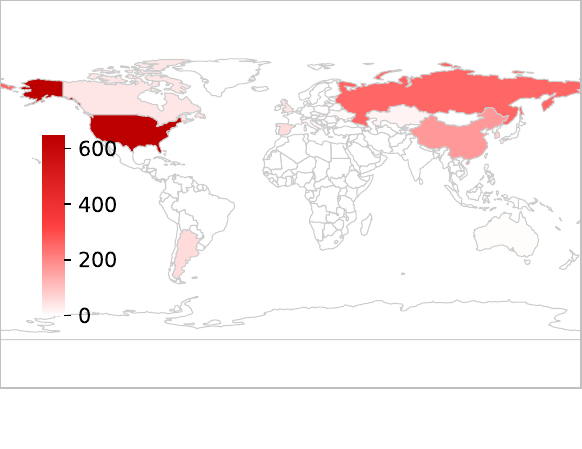}}
        \label{fig:map}%
    }
    \caption{Analysis and visualization of \SystemName\ dataset.}
    \label{fig:vis}
\end{figure*}

\vspace{-4pt}
\section{\SystemName~Analysis and Results}
\subsection{RQ1. Analysis and Results}
\subsubsection{Platforms}

As depicted in Table \ref{tab:sum}, \textbf{YouTube} stands as the largest video-sharing platform, hosting the highest amount of deepfake content. One of YouTube's distinguishing features is that its system allows uploads of up to 15 minutes, in contrast to TikTok's limit of 3 minutes. Among the 2,000 \SystemName~videos, 277 videos exceed the 3-minute threshold, with YouTube leading with 247 uploads. Moreover, YouTube has shown significant growth in deepfake videos each year compared to the previous year. The increase in 2019 is from 9 to 137, representing a 1,422\% increase (Fig.~\ref{fig:when}).

Also, \textbf{TikTok} has experienced remarkable growth,
accompanied by a substantial rise in deepfake videos over the years. 
Despite having fewer uploads (599 videos), TikTok outperforms YouTube (3,187,806 likes and 264,270 comments) in engagement with over 41,357,828 likes and 550,348 comments.
Interestingly, approximately 45\% of deepfake videos on TikTok feature dance-related content with famous individuals, gaining popularity among users, especially the younger demographic. This can be attributed to the nature of TikTok, known for its short videos averaging around 20 seconds in length. Moreover, mobile apps like Faceplay facilitate the creation of these short videos, leading to deepfake trends on TikTok. 
On the other hand, \textbf{Bilibili} has witnessed fluctuations in its growth over the years, with varying numbers of uploaded videos.
However, intriguing observations are made on the Bilibili platform. 
While we are unable to find deepfake videos uploaded by Chinese individuals on the YouTube, platforms developed by Chinese companies, such as TikTok and Bilibili, have emerged as sources for Chinese deepfake videos.
Unlike TikTok, predominantly Chinese creators upload deepfake videos on the Bilibili. 

This data proves to be valuable in addressing national and racial bias, considering the majority of deepfakes are created in the United States (US) and the majority of victims are Caucasians. 
Additionally, we find that deepfake videos related to Chinese politicians are blocked and inaccessible through regular searches. This indicates a level of censorship and control over deepfake content involving political figures in China.

\textbf{Reddit} has shown a relatively modest growth in uploaded videos. However, Reddit's unique strength lies in fostering communities where users engage in sharing and discussing content. During our data collection, we identified 14 user accounts across three exclusive deepfake communities solely present on Reddit.
Remarkably, out of the 111 deepfake videos found on Reddit, a significant 46\% (52 videos) are of high resolution, exceeding 1,000×1,000 pixels. This proportion stands out considering that only around 5\% (91 videos) of the entire dataset achieved such a level of clarity. This observation underscores the significant role Reddit plays in cultivating and circulating a noteworthy quantity of high-quality deepfake videos within its dedicated communities.

\vspace{-5pt}
\subsubsection{Generation Methods} 
We identify a total of 16 popular deepfake generation methods used as illustrated in Fig.~\ref{fig:prop_model}, and we group them into three main groups: open source, mobile application, and commercial provider. 

\textbf{Open source. } 
These frameworks are accessible to semi-professionals aiming to create high-quality deepfake videos. Notable examples encompass DeepFaceLab~\cite{DeepFaceLab}, 
DeepFaceLive~\cite{DeepfaceLive}, First-Order-Motion~\cite{siarohin2019first}, and SimSwap~\cite{Chen_2020}.  Particularly, DeepFaceLab finds extensive use for tasks like face replacement, expression conversion, voice synthesis, and de-aging due to its high resolution and fast image processing speed. As a result, DeepFaceLab contributes to 74.8\% of all usage methods, as depicted in Fig. \ref{fig:prop_model}.  

\textbf{Mobile app. } 
Mobile apps provide user-friendly interfaces for easy deepfake video creation. They offer convenience with shorter processing times compared to other methods. Examples include FacePlay~\cite{FacePlay}, Reface~\cite{reface}, Deepfake Studio~\cite{DeepfakeStudio}, FaceApp~\cite{FaceApp}, Revive ~\cite{Revive}, LicoLico~\cite{licolico}, Fakeit~\cite{Fakeit}, and DeepFaker~\cite{DeepFaker}. 

Among videos created by these apps, 20\% are under 30 seconds, with over half made using FacePlay due to its creative filters and animations. Notably, this dominance is driven by just two FacePlay uploaders, showcasing concentrated content creation by a single user.

\textbf{Commercial product. } Regarding commercial deepfake providers, our dataset includes DeepFakesWeb~\cite{DeepFakesWeb}, Deepcake.io~\cite{Deepcakeio}, DeepFaker Bot~\cite{DeepFakerBot}, and Revel.ai~\cite{Revelai}. These companies offer professional services and advanced techniques for realistic results.
We observe variations in AI studio choices by uploaders based on their countries. North American uploaders (US, Canada) favor Revel.ai and DeepFakerWeb (70\% of Deepfake AI Studio videos). Russian uploaders prefer Deepcake.io and DeepFaker Bot (30\% of videos), indicating a prevalence of American-affiliated studios even within Deepfake AI Studio.

\subsubsection{Creators/Uploaders} 

Among the 2,000 videos, the origin country of the uploaders could be determined in 1,510 videos.
The majority of the videos are generated in the US, as shown in Fig.~\ref{fig:map}. Out of the total 1,510 videos, 647 videos are distributed by US accounts, accounting for 42.8\% 
of the total. Russia and China account for  259 videos (17.1\%) 
and 174 videos (11.5\%), 
respectively. 
Other countries like Korea, Argentina, Spain, and others have less than 100 videos distributed.

We identified details about 235 uploaders out of 258 (excluding unknown data) from 21 nations. Russia leads with 49.8\% (117 individuals), followed by China at 23.0\% (54 individuals), and the US at 10.2\%. This suggests heavy uploading by US accounts, whereas Russian and Chinese uploaders are generally lighter users. Heavy uploaders usually maintain dedicated channels for entertainment. Analyzing political uploads, individuals contribute less than 6 videos on average, below the overall average of 7.5 videos per uploader, potentially due to account locks or the use of multiple accounts.
 
\subsection{RQ2. Analysis and Results}
\subsubsection{Deepfake Victims/Drivers} 

We illustrate the gender distribution across ethnicities in Fig.~\ref{fig:eth_gender_dist}. Deepfakes demonstrate variations based on race, skin color, and gender. For the \textbf{race} category, Victim and driver data are categorized into Caucasian, Asian (East), Asian (South), Asian (Middle East), Asian (South East), and African (Black). For victims, Caucasian cases total 1,463, followed by Asian (East) with 251, and African (Black) with 58. Driver counts are 1,297 Caucasian, 162 Asian (East), and 22 Asian (South). \textbf{Skin color} analysis reveals 1,463 White victims, 402 Brown, and 79 Black. Among drivers, 1,240 are White, 211 are Brown, and 80 are Black. \textbf{Gender} distribution shows 1,528 male victims and 445 females, a difference of about 3.4 times. For drivers, the counts are 1,297 males and 402 females, indicating over three times more male targets in both victim and driver roles for deepfakes.
 
\subsubsection{Intentions} With a focus on the creation of deepfakes for more malicious purposes, we conducted a classification process. First, data linked to intentions such as financial gain or tarnishing the reputation of specific individuals were categorized as Fraud. Secondly, content containing political implications or portraying politicians was classified as Politic, and content used for entertainment purposes rather than malicious intentions was categorized as Entertainment. This differentiation allows us to discern the underlying motivations of deepfake creators. While a single video may encompass multiple intentions, we prioritized categorization based on the most predominant intent for accurate statistical analysis. Upon such differentiation, we observed a consistent increase in the volume of deepfakes across all mentioned categories, as shown in Fig. ~\ref{fig:purpose}.

\begin{figure}[!t]
\vspace{-10pt}
    \centering
        \subfloat[Distribution of victim and driver gender across ethnicities
    ]{%
      {\includegraphics[width=5.6cm]{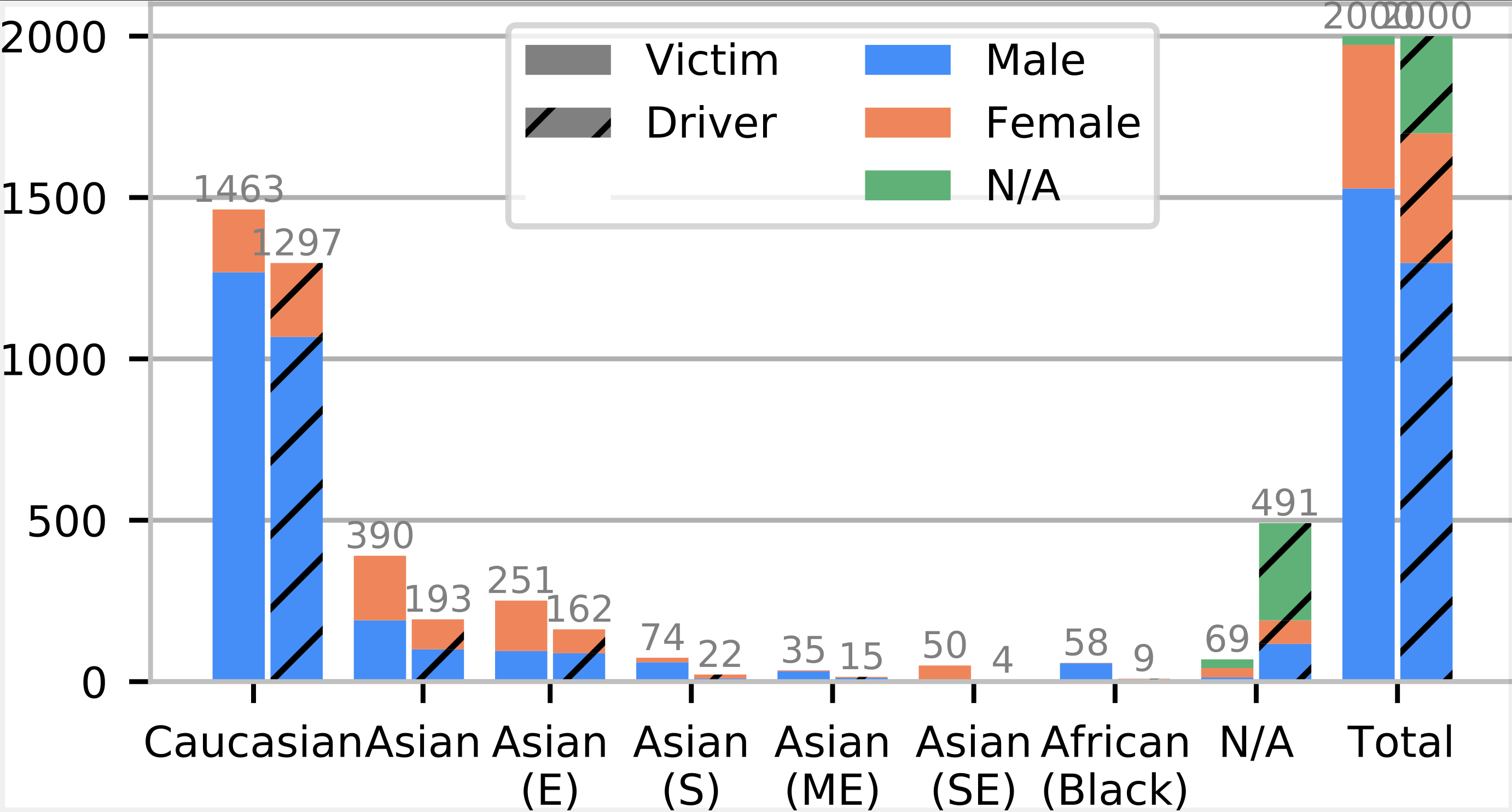}}
        \label{fig:eth_gender_dist}%
    }\hspace{2pt}
    \subfloat[Purpose of deepfake generation]{%
       {\includegraphics[width=2.4cm]{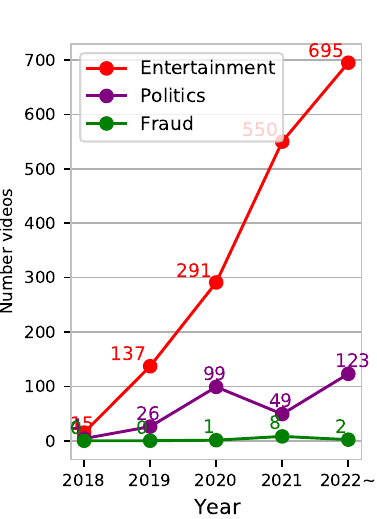}}
        \label{fig:purpose}%
    }
   \caption{Analysis on targets of deepfakes.}
\label{fig:vis_analysis}
\end{figure}

\textbf{Entertainment deepfakes.} The prevalence of entertainment-based deepfake videos is evident across all classifications, exhibiting a remarkable annual growth rate of over 100\%. This dominance can be attributed to various factors, including the abundance of victim data and ease of generation in the entertainment domain. Additionally, leveraging the fame and popularity of well-known individuals allows these deepfakes to capture attention effectively, as seen in parodies of movies and dance videos featuring popular singers. 

\textbf{Political and fraud deepfakes.}  In contrast, political deepfake videos, as shown in Fig. \ref{fig:purpose},  demonstrate a lower quantity compared to entertainment, but they exhibit a growing trend. The number of political deepfakes created for political purposes has significantly increased from 4 videos in 2018 to 123 videos in 2022, indicating a staggering growth of 2,975\% over four years. 
These political deepfakes primarily target influential figures such as presidents, prime ministers, and kings, accounting for approximately 70\% (300 videos) of the total 428 deepfake videos.

Furthermore, deepfake technology has been maliciously employed for fraudulent purposes. Notable instances include videos featuring famous celebrities or business figures endorsing specific companies or products to deceive consumers.For example, the attractiveness of American actor Leonardo DiCaprio is exploited to promote Russian amusement parks and products, or the image of American entrepreneur Elon Musk is utilized to endorse Chinese products, revealing instances of deceptive content.

\textbf{Note. } This is evident in the significant increase (381\%) of political deepfake videos during the 2020 US presidential election, with 53\% (53 videos) of the total 99 political deepfakes specifically targeting candidates Donald Trump and Joe Biden, illustrating a distinct trend of political manipulation. Additionally, malicious exploitation of deepfakes can be observed, such as the creation of offensive videos featuring Korean celebrities to defame their reputations. These deceitful actions by scammers and cybercriminals pose serious risks, including identity theft, financial fraud, and various illegal activities in the future. 


 \begin{figure}[!t]
    \centering

    \includegraphics[width=7.4cm]{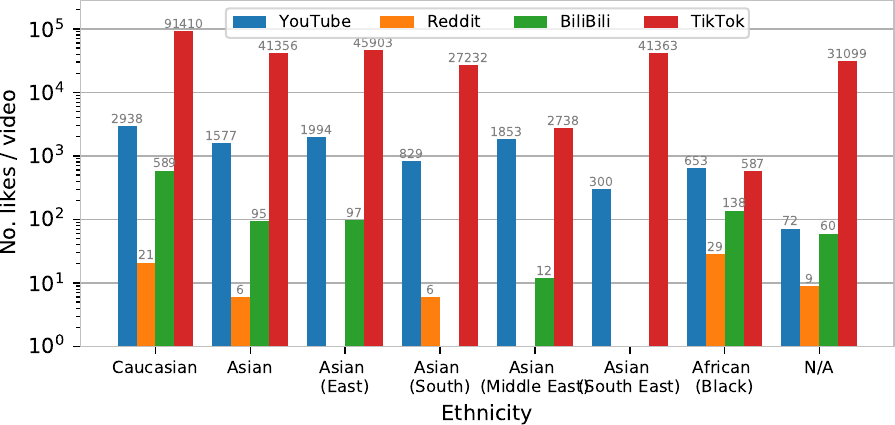}
    \vspace{-8pt}    \caption{Distribution of the average number of likes per video over different victim ethnicities. } 
    \label{fig:avg_inter2}

\end{figure}

\subsection{RQ3. Analysis and Results}

\subsubsection{Viewer's Interest towards Victim's Ethnicity} 
First, our research seeks to understand the relationship between different victim ethnicities and the level of interaction they receive from subscribers on various platforms. In Fig. \ref{fig:avg_inter2}, we calculate the average number of likes across platforms and ethnicities. It's crucial to comprehend which platforms facilitate greater viewer interaction and the specific demographics they engage with. Consequently, future efforts in detecting deepfakes can allocate more resources towards these identified tendencies. {In terms of platform comparison, TikTok is emerging as a hub to attract views, as demonstrated by its highest number of likes per video across all ethnicities. This trend could potentially encourage cybercriminals to propagate deepfake content more on this platform compared to others. From an ethnicity perspective, while Caucasian subjects dominate in the number of videos they are featured in, as shown in Fig. \ref{fig:eth_gender_dist}, they also garner the most interest from viewers, demonstrated through their high levels of interaction across platforms, which can be up to 91,410 likes per video on TikTok. Meanwhile, the interaction with videos featuring African (Black) subjects is considerably lower. 
}
 \begin{figure}[!t]
    \vspace{0pt}
    \centering
    \includegraphics[width=7.9cm]{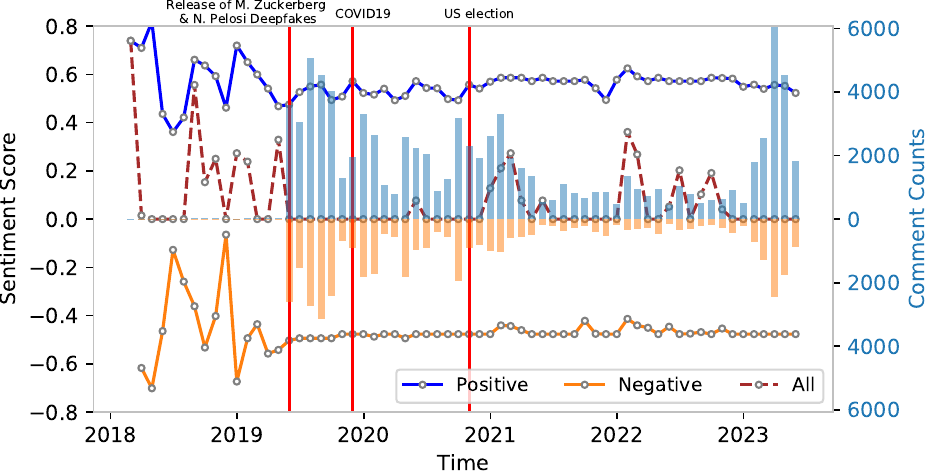}
    \caption{Temporal changes in video comment counts and their corresponding median sentiment scores, with red solid vertical lines indicating significant events.} \label{fig:sent_time}
\end{figure}
\subsubsection{Viewer's Response over Time}
\label{subsec:sentiment_time}
Here, we explore deeper into the viewers' attitudes toward the videos over time. We note that YouTube videos make up a significant part of our dataset, and it also provides a user-friendly API for collecting comments along with their timestamps. We decide to conduct our sentiment analysis on our YouTube videos, without compromising the generalizability of our observations. We use the pre-trained VADER \cite{hutto2014vader}, 
a lexicon and rule-based sentiment analysis tool designed to express sentiment polarity and intensity. 
{Our findings are illustrated in Fig. \ref{fig:sent_time}. At first glance, people most of the time are divided into two equal groups with different attitudes towards deepfake, as indicated by the overall median sentiment score, which mostly lies at 0. 
It is evident that right after deepfakes of public figures, such as Mark Zuckerberg \cite{RachelCNN2019} and Nancy Pelosi \cite{DonieCNN2019}, were posted on social media in 2019, it sparked wide-ranging debates  about their potential misuse. Before, during, and after the pandemic, as well as during the US presidential election, deepfakes continued to spread widely, as demonstrated by a high volume of comments. This was a time when people became more concerned about their online appearance and the potential significant impact deepfakes could have on their lives. However, from mid-2021 onwards, interest in deepfake videos began to wane, and comments became more positive.}
{In early 2023, viewer interest in deepfake videos rebounded, reaching  an average of 6,000 comments per video in March, as evidenced by an increase in comments. This phenomenon could be partially attributed to the emergence of powerful generative models, such as MidJourney \cite{midjourney}, which can produce highly realistic deepfakes.
}
\begin{figure}[!t]
    \vspace{0pt}
    \centering
    \includegraphics[width=7.9cm]{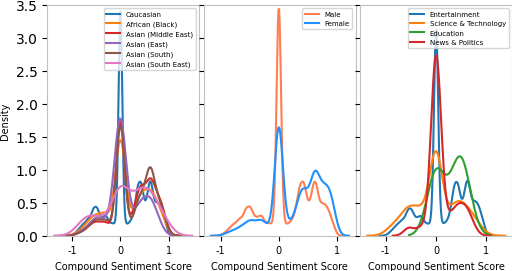}   
    \caption{Sentiment score distribution. The higher the score, the more positive viewers' attitudes are towards the videos.}
\label{fig:sent_dist_all}
\end{figure} 
\subsubsection{Viewer's Attitudes in Various Categories}
Finally, we analyze the sentiment score distribution concerning various factors: the ethnicity and gender of the victim, vs. video categories. We anticipate that viewers' attitudes toward different aspects of video segmentation can reveal their biases. Consequently, we expect that by implementing appropriate countermeasures targeting specific areas, the propagation of deepfakes across platforms can be hindered. We employ the same settings as in Sec. \ref{subsec:sentiment_time} and provide the results in Fig. \ref{fig:sent_dist_all}. {In terms of ethnicity, we can observe that videos involving Caucasian victims, who are dominant in our dataset (Fig. \ref{fig:eth_gender_dist}), receive the most neutral comments compared to other ethnicities. Furthermore, videos involving victims from Southeast Asia, with a smaller number of videos, elicit relatively positive comments, as shown by their distribution being mostly greater than 0.} 

When considering the gender of the victim, we observed that female victims tend to receive more positive comments from viewers. Furthermore, we conducted a Mann-Whitney U test to compare the means of two distributions \cite{mann1947test} with  $\alpha$ set at 0.01. The obtained p-value  was extremely small (p < 0.001), providing strong evidence that the mean sentiment scores of female victims are significantly higher than those of male victims. This observation may be partially attributed to the fact that female victims are typically cast in entertainment videos, such as those featuring dancing or singing, which may evoke less concern about their potential harm.
Lastly, we classify the sentiment scores according to different video categories, which are specified by creators when uploading their videos on YouTube. We observe that videos labeled as educational content receive more positive comments, revealing a potential flaw that attackers might exploit to widely spread deepfake content. Conversely, Science and Technology videos, which are typically aimed at educating people about deepfakes, receive more negative comments compared to other categories. Typical negative comments in this group are: ``This tech is so dangerous'' or ``This is scary''.

\section{Conclusion and Discussion}
In this work, we thoroughly investigate deepfake videos across various platforms and introduce a newly collected deepfake dataset, \SystemName{}. Additionally, we conduct insightful analyses on different aspects of the data, considering the perspectives of both creators and viewers. Through this work, we uncover previously unexplored observations, such as the bias of creating tools and among high-frequency uploaders, as well as the diverse attitudes of viewers toward victims of varying ethnicities across different video categories.
We believe that our findings can contribute to future research by enabling the development of more robust and effective deepfake detection techniques in real-world scenarios, which are often challenging and unpredictable from various perspectives. \\

\noindent\small{ \textbf{Acknowledgement.}
This work was partly supported by Institute for Information \& communication Technology Planning \& evaluation (IITP) grants funded by the Korean government MSIT: (No. 2022-0-01199, Graduate School of Convergence Security at Sungkyunkwan University), (No. 2022-0-01045, Self-directed Multi-Modal Intelligence for solving unknown, open domain problems), (No. 2022-0-00688, AI Platform to Fully Adapt and Reflect Privacy-Policy Changes), (No. 2021-0-02068, Artificial Intelligence Innovation Hub), (No. 2019-0-00421, AI Graduate School Support Program at Sungkyunkwan University), and (No. RS-2023-00230337, Advanced and Proactive AI Platform Research and Development Against Malicious deepfakes).}

{\small
\bibliographystyle{ieee_fullname}
\bibliography{bib}
}

\end{document}